# DISCUSSION OF: STATISTICAL ANALYSIS OF AN ARCHEOLOGICAL FIND

By Stephen M. Stigler

*University of Chicago*

Statistics is the science of uncertainty, and it should be capable of helping to address even hard to quantify problems. Indeed, the very attempt to quantify may itself shed light and understanding, and can often lead to better articulation of even qualitative evidential arguments. Yet, when statistical ideas are used in areas where wide segments of the population hold strongly divided passionate views, areas such religion or politics, the entry of statistics into the discussion is seldom accorded a warm and friendly reception. Instead, the greeting is at best extraordinarily skeptical, with quibbling over minor points that would be passed by silently in less-contentious studies, and with inhospitality to even the best of intentions. At worst, the intruder is burned at the stake or removed from the rolls of the employed, although such extremes are rarer these days than they were at the time of Giordano Bruno and Galileo.

Is this resistance rational? Do questions like that confronted in Andrey Feuerverger's painstakingly honest study of an archeological find, questions involving broad public knowledge and wide publicity, require a different standard of proof than run-of-the mill scientific questions? I think they may well, for several reasons.

1. The very wide public attention to the area, even before the discovery of the evidence, changes the way we think of the evidence. For example, the temptations to persons of unknown identity (even in the distant past) to fraudulently manufacture evidence must be considered, and the weighing of potential forms of fraud in any modeling context is a highly vexing question.

2. Even aside from any possible fraud, the conditions surrounding the arrival of the evidence can legitimately raise questions that would never arise in more mundane investigations. For example, we are told that, "No information is available regarding the placement of the various ossuaries among the *kokhim*." But the names involved in this case are so universally recognized that it might be argued that the absence of information is in this case informative, as the dog who did not bark was to Sherlock Holmes. One might believe that had the ossuaries been arrayed together in a meaningful order, this would with some probability have been noted, and the lack of such notation suggests they were not.

3. Francis Galton issued a caution in 1863 for those dealing with small data sets with uncertain generating mechanisms: "Exercising the right of occasional suppression and slight modification, it is truly absurd to see how plastic a limited number of observations become, in the hands of men with preconceived ideas" (*Meteorographica*, London: Macmillan, 1863, page 5). Since occasional suppression and slight modification can be a part of sound statistical analysis, it is easy to overlook this potential bias, for it will not always be obviously present or consciously operating in a deceptive way.

I commend Andrey Feuerverger for undertaking this investigation. That it may be greeted skeptically is no reflection upon him, only upon the nature of the question he considers. Some of the assumptions he forthrightly makes, such as the independent assignment of names in families, may not survive later scrutiny. But in the face of all these difficulties, his carefully qualified analysis reminds us that addressing a question is not the same as resolving it, and that issues of wide general interest where prior opinions are sharply divided present novel problems of statistical formulation. I look forward to the ensuing dialogue, which will hopefully have greater focus because of the pains Feuerverger has taken to frame and present the issues.

Department of Statistics
University of Chicago
5734 University Avenue
Chicago, Illinois 60637
USA
E-mail: stigler@uchicago.edu